\documentclass[showpacs,preprintnumbers,
superscriptaddress,amsmath]{revtex4}

\usepackage{epsfig}

\newcommand{\PLB}[3]{Phys.\ Lett.\ B\ {\bf #1},\ #2 (#3)}

\newcommand{\PRD}[3]{Phys.\ Rev.\ D\ {\bf #1},\ #2 (#3)}

\newcommand\e{\epsilon}

\newcommand\rr{\rho}

\newcommand{\be}{\begin{equation}}
\newcommand{\ee}{\end{equation}}
\newcommand{\bea}{\begin{eqnarray}}
\newcommand{\eea}{\end{eqnarray}}
\newcommand{\ba}[1]{\begin{array}{#1}}
\newcommand{\ea}{\end{array}}

\begin{document}

\title{Gluon self-energy in the color-flavor-locked phase}

\author{H.\ Malekzadeh}
\email{malekzadeh@figss.uni-frankfurt.de}
\affiliation{Frankfurt International Graduate School of Science, 
J.W.\ Goethe-Universit\"at, Max von Laue-Str.\ 1,
D-60438 Frankfurt am Main, Germany}

\author{Dirk H.\ Rischke}
\email{drischke@th.physik.uni-frankfurt.de}
\affiliation{Institut f\"ur Theoretische Physik and
Frankfurt Institute for Advanced Studies,
J.W.\ Goethe-Universit\"at, Max von Laue-Str.\ 1,
D-60438 Frankfurt am Main, Germany}

\date{\today}

\begin{abstract}
We calculate the self-energies and the spectral densities of 
longitudinal and transverse gluons
at zero temperature in color-superconducting quark matter in
the color-flavor-locked (CFL) phase.
We find a collective excitation, a plasmon, at energies smaller
than two times the gap parameter and momenta smaller than about
eight times the gap. The dispersion relation of this mode
exhibits a minimum at some nonzero value of momentum, indicating
a van Hove singularity.
\end{abstract}

\maketitle

At asymptotically large quark chemical potentials $\mu$ 
and sufficiently small temperatures $T$,
quark matter is a color superconductor \cite{bailinlove}.
While there are, in principle, many different color-superconducting
phases, corresponding to the different possibilities to form
quark Cooper pairs with definite color, flavor, and spin quantum numbers,
for three quark flavors and quark chemical potentials
much larger than the strange quark mass, the ground state of
color-superconducting quark matter is the
so-called color-flavor-locked (CFL) phase \cite{arw}.

In the CFL phase, the $SU(3)_c \times SU(3)_f$ color and (vector)
flavor symmetry of QCD is broken to the diagonal subgroup
$SU(3)_{c+f}$. Consequently, all eight gluons become massive due
to the Anderson-Higgs mechanism.
The situation is quite similar to that in the 2SC case, although
there $SU(3)_c$ is broken to $SU(2)_c$, and
only five gluons become massive, while the three gluons
of the residual $SU(2)_c$ color symmetry remain massless.
These general symmetry considerations can be confirmed by
an explicit calculation of the gluon Meissner masses in a
given color-superconducting phase.
The Meissner mass (squared) of a gluon with adjoint color $a$
is the zero-momentum limit of the transverse component of the
gluon self-energy at zero energy, 
$\lim_{p \rightarrow 0} \Pi^t_{aa}(p_0=0,{\bf p})$.

At asymptotically large $\mu$, the QCD coupling constant
$g \ll 1$, and the gluon self-energy is dominated 
by the contributions from one quark and one gluon loop.
The quark loop is $\sim g^2 \mu^2$, while the gluon loops
are $\sim g^2 T^2$.
Since the color-superconducting
gap parameter is $\phi \sim \mu \exp(-1/g) \ll \mu$ 
\cite{son}, and since the transition temperature to the
normal conducting phase is $T_c \sim  \phi$,
for temperatures where quark matter
is in the color-superconducting phase, $T \lesssim T_c \ll \mu$,
the gluon loop contribution can be neglected.
Following this line of arguments, the gluon Meissner masses have been computed 
for the 2SC phase in Ref.\ \cite{meissner2} and for the CFL phase 
in Ref.\ \cite{meissner3,sonstephanov}.
The full energy-momentum dependence of the one-loop gluon self-energy 
has also been computed, but so far only for the 2SC phase \cite{2cf,dirkigor}.
The corresponding calculation for the CFL phase 
is the goal of the present paper. 

For the 2SC phase,
the derivation of the gluon self-energies and
propagators was explicitly discussed in Sec.\ II of
Ref.\ \cite{dirkigor}, and we briefly remind the reader
of the main steps.
Just like in other systems (ordinary superconductors, the standard
model of electroweak interactions) where gauge symmetries are broken,
there are Nambu-Goldstone fluctuations which
mix with the longitudinal components of the gauge field. 
For a particular choice of gauge ('t Hooft gauge), 
the Nambu-Goldstone modes decouple 
from the longitudinal components of the
gauge field. Furthermore, for a particular choice of the 
't Hooft gauge parameter, 
the Nambu-Goldstone modes can be eliminated and the
gauge field propagator is explicitly 4-transverse.

In our case, i.e., for the CFL phase, 
the computational steps leading to the gluon
self-energies $\Pi^{\mu \nu}_{ab}$ and the gluon propagators
$\Delta^{\mu \nu}_{ab}$
are quite similar to those in the 2SC phase \cite{dirkigor}, 
if not even simpler because all eight gluons are
affected similarly by the breaking of the gauge symmetry,
\be 
\Pi^{\mu \nu}_{ab} (P) \equiv \delta_{ab} \, \Pi^{\mu\nu}(P)\;, \;\;\;
\Delta^{\mu \nu}_{ab} (P) \equiv \delta_{ab} \, \Delta^{\mu\nu}(P)\;,
\;\;\; a,b = 1, \ldots, 8\;.
\ee
Therefore, we do not give the details of 
this lengthy but straightforward calculation and just
quote the final result.
The final answer for the propagator of transverse gluons reads
[cf.\ Eq.\ (56) of Ref.\ \cite{dirkigor}]
\be
\Delta^t (P) = \frac{1}{P^2 - \Pi^t(P)}\;,
\ee
while for longitudinal gluons we have [cf.\ Eq.\ (57) of
Ref.\ \cite{dirkigor}]
\be
\hat{\Delta}^{00} (P) = - \frac{1}{p^2 - \hat{\Pi}^{00}(P)}\;.
\ee
Here, the transverse and longitudinal components of the
gluon self-energy are computed from projections of $\Pi^{\mu \nu}$,
\begin{subequations}
\bea
\Pi^t(P) & = &\frac{1}{2} \left( \delta^{ij} - \hat{p}^i \hat{p}^j
\right) \Pi^{ij}(P)\;,\\
\hat{\Pi}^{00}(P) & = &  p^2 \,
\frac{\Pi^{00}(P)\,\Pi^\ell(P) - 
\left[ \Pi^{0i}(P) \, \hat{p}_i \right]^2 }{
p_0^2 \, \Pi^{00}(P) + 2\, p_0\,p\, \Pi^{0i}(P) \, \hat{p}_i
+ p^2 \, \Pi^\ell(P) } \;, \label{hatPi}
\eea
where
\be
\Pi^\ell(P) = \hat{p}^i\, \Pi^{ij}(P)\, \hat{p}^j\;.
\ee
\end{subequations}
The computation of the individual components and projections 
$\Pi^{00}, \, \Pi^{0i},\, \Pi^t,$ and $\Pi^\ell$ is lengthy and thus
deferred to the appendix. Note that the particular form of the longitudinal
self-energy $\hat{\Pi}^{00}$ arises from decoupling spatially longitudinal and
time-like gluon degrees of freedom, see Ref.\ \cite{dirkigor}.

\begin{figure}
\begin{center}
\vspace*{1cm}
\includegraphics[width=15cm]{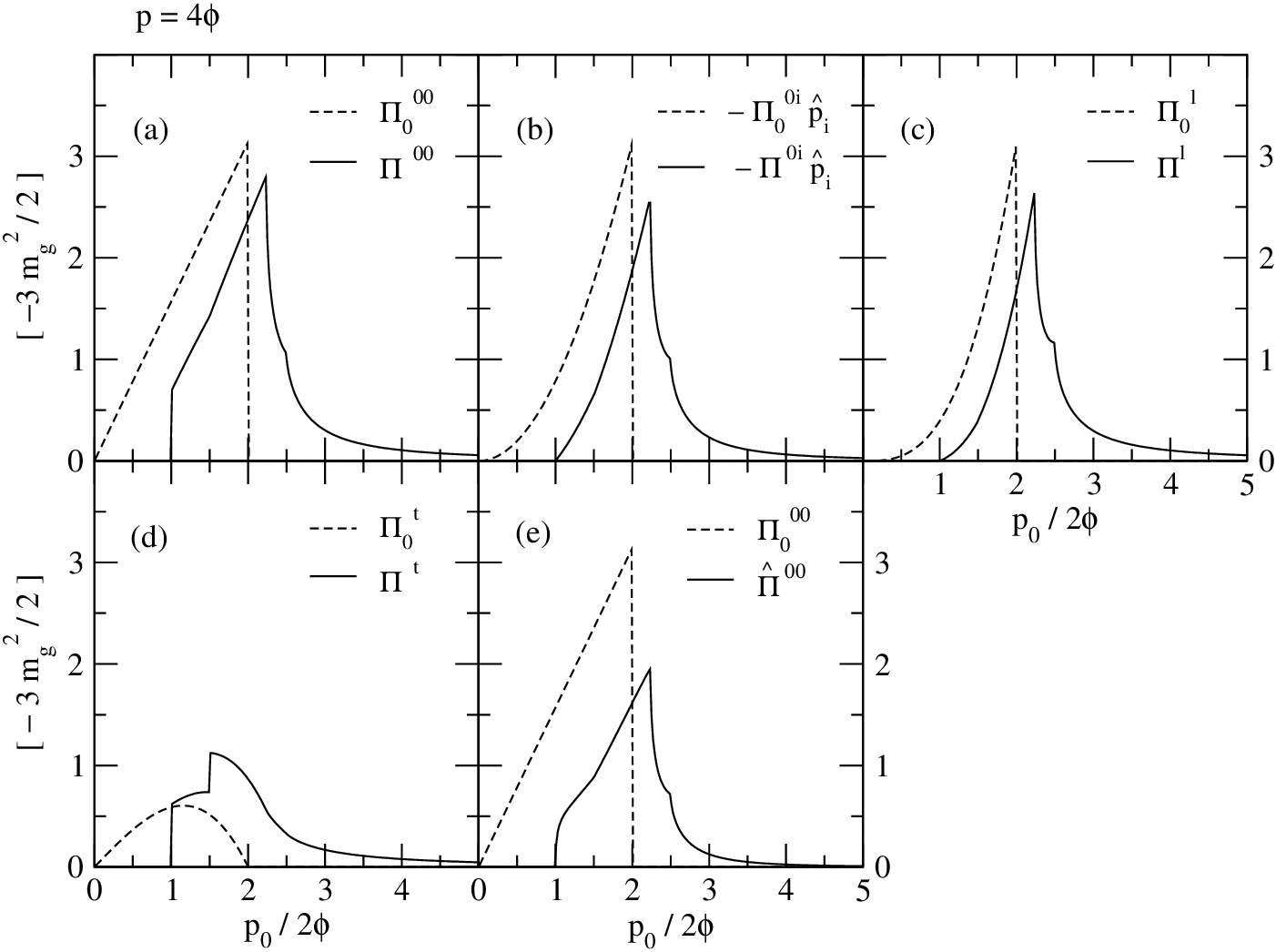}
\caption
{The imaginary parts of (a) $\Pi^{00}$, (b) $- \Pi^{0i} \hat{p}_i$,
(c) $\Pi^\ell$, (d) $\Pi^t$, and (e) $\hat{\Pi}^{00}$
as a function of energy $p_{0}$ for a gluon momentum $p=4\phi$. 
The solid lines are for the CFL phase, the dotted lines correspond
to the HDL self-energy.}
\label{im00}
\end{center}
\end{figure}

In Fig.\ \ref{im00} we show the imaginary part of several
components of the gluon self-energy for a gluon momentum $p = 4\phi$ 
as a function of the gluon energy $p_{0}$. For comparison, we
also show the corresponding results for the gluon self-energy
in the ``hard-dense loop'' (HDL) limit, $\Pi^{\mu \nu}_0$,
cf.\ Eqs.\ (63a), (65a), (69a) and (69b) of Ref.\ \cite{dirkigor}.
The imaginary parts are quite similar to those of the 2SC case, 
cf.\ Fig.\ 1 of Ref.\ \cite{dirkigor}.
There are subtle differences, though, due to
appearance of two kinds of gapped quark excitations, one so-called singlet
excitation with a gap $\phi_{\bf 1}$, and eight so-called
octet excitations with a gap $\phi_{\bf 8} \equiv \phi$ \cite{arw}.
In weak coupling, the singlet gap is approximately twice as large
as the octet gap, $\phi_{\bf 1} \simeq 2\, \phi_{\bf 8} \equiv 2 \, \phi$
\cite{tomschafer1,igor}. 
The one-loop gluon self-energy in the CFL phase has two types of
contributions, depending on whether the quarks in the loop
correspond to singlet or octet excitations, 
cf.\ Eq.\ (23b) of Ref.\ \cite{meissner3}. 
For the first type, both quarks in the loop
are octet excitations, and for the second, one is an octet and
the other a singlet excitation. 
There is no contribution from singlet-singlet excitations.

Nonvanishing octet-octet excitations require gluon
energies to be larger than $2\, \phi_{\bf 8} \equiv 2\, \phi$, 
while octet-singlet excitations
require a larger gluon energy, $p_0 \geq \phi_{\bf 1} + \phi_{\bf 8}
\equiv 3\, \phi$.
This introduces some additional structure in the imaginary parts
at $p_0 = 3 \phi$, which can be seen particularly well in Figs.\ \ref{im00}
(d) and (e). 

Some imaginary parts exhibit a peak-like structure at
a gluon energy $p_0 = E_{p}^{\bf 88} \equiv 
\sqrt{p^2 + (\phi_{\bf 8}+ \phi_{\bf 8})^2} = \sqrt{20}\,
\phi$, followed by a sharp drop for larger energies.
The mathematical reason is seen in Eqs.\ (\ref{A11}) and (\ref{A13}), where
the regions $p_0 > E_{p}^{\bf 88}$ and $p_0 \leq  E_{p}^{\bf 88}$ are
separated by $\Theta$ functions.
In the normal phase, the imaginary parts of the gluon self-energies 
actually vanish above $p_0 = p$. In color-superconducting phases, 
the imaginary parts do not vanish but fall off rapidly. This has
already been noted for the 2SC phase \cite{dirkigor}, and is
confirmed here by the results for the CFL phase.
If there is a peak-like structure at $p_0 = E_p^{\bf 88}$,
from Eqs.\ (\ref{A11}) and (\ref{A13}) we expect a similar peak to appear at
$p_0 = E_p^{\bf 1 8} \equiv \sqrt{p^2 + 
(\phi_{\bf 1} +\phi_{\bf 8})^2} = 5\, \phi$. One indeed
sees an additional structure at this point, but it
is much less pronounced since it is
located on top of the sharp drop of the first peak.

\begin{figure}[ht]
\begin{center}
\vspace*{1cm}
\includegraphics[width=15cm]{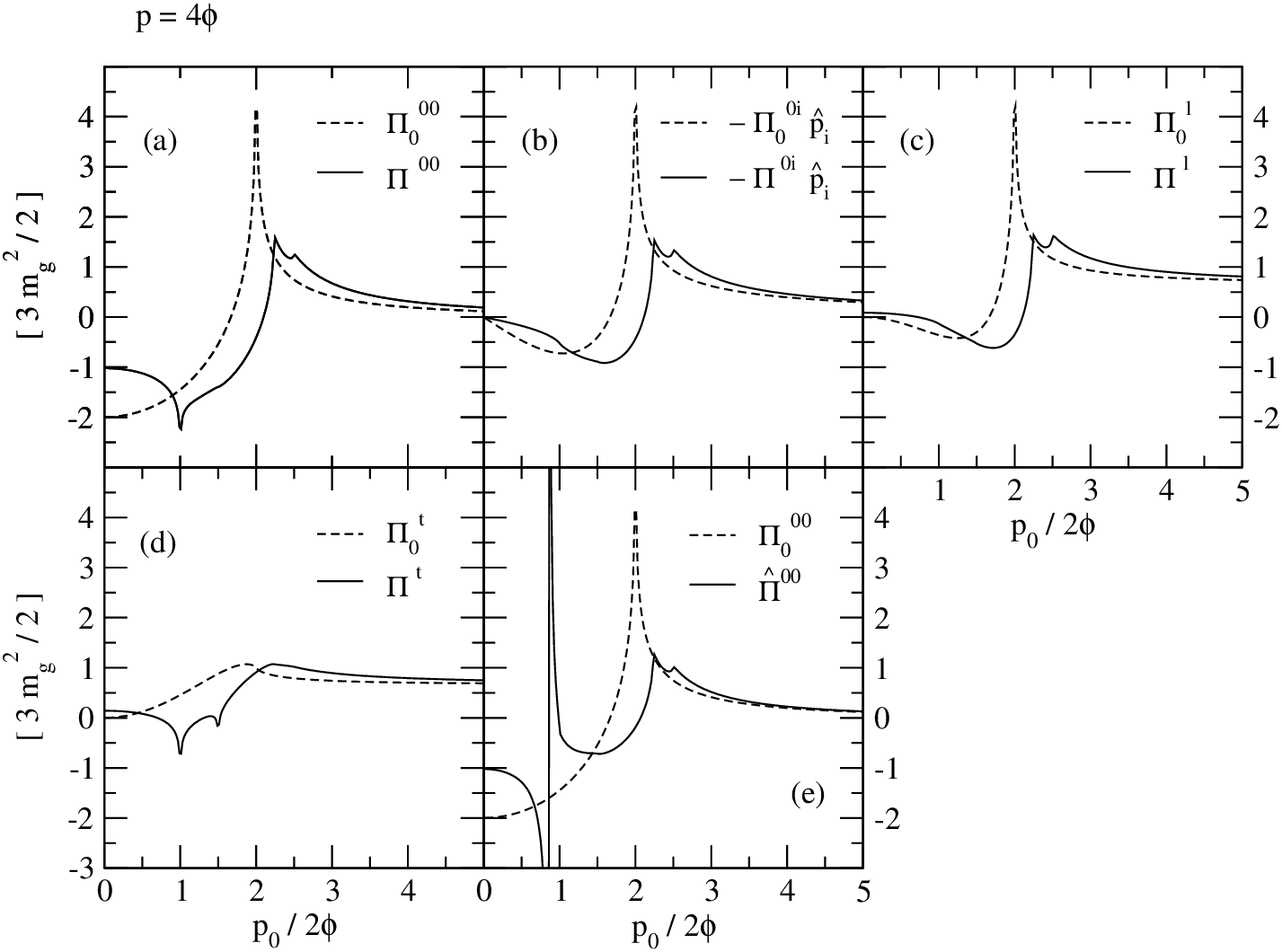}
\caption
{The real parts of (a) $\Pi^{00}$, (b) $- \Pi^{0i} \hat{p}_i$,
(c) $\Pi^\ell$, (d) $\Pi^t$, and (e) $\hat{\Pi}^{00}$
as a function of energy $p_{0}$ for a gluon momentum $p=4\phi$. 
The solid lines are for the CFL phase, the dotted lines correspond
to the HDL self-energy.}
\label{re}
\end{center}
\end{figure}

In Fig.\ \ref{re} we show the real parts of the gluon self-energy
corresponding to the imaginary parts shown in Fig.\ \ref{im00}.
These are quite similar to the ones in the 2SC phase, cf.\ Fig.\ 2
of Ref.\ \cite{dirkigor}. An explanation of the various structures
can be given following arguments similar to those of Refs.\ 
\cite{2cf,dirkigor}. In essence, when computing the real
part from a dispersion integral over the imaginary part, cf.\
Eq.\ (\ref{disp}), a change of gradient in the imaginary part
leads to a cusp-like structure in the real part. The only 
function that does not fit this general rule is 
${\rm Re}\,\hat{\Pi}^{00}$. The reason is that this function is computed
from the real part of the right-hand side of Eq.\ (\ref{hatPi}).
Note the singularity at $p_0 \simeq 1.6\, \phi$. 
This singularity is caused by a root of the denominator in
Eq.\ (\ref{hatPi}), $P_\mu \, \Pi^{\mu\nu}(P) \, P_\nu =0$.
As shown in Ref.\ \cite{zarembo} (see also Ref.\ \cite{dirkigor}
for the 2SC case),
this condition determines the dispersion relation of the 
Goldstone excitations.
As one expects, for large energies $p_{0}\gg\phi$ 
the real parts of the self-energies 
approach the corresponding HDL limit. Deviations from the HDL limit 
occur only for gluon energies $p_{0}\sim\phi$.  

We now compute the spectral densities from
the real and imaginary parts of the gluon self-energies.
When ${\rm Im}\, \hat{\Pi}^{00}(p_{0},{\bf p}),
{\rm Im}\, \Pi^{t}(p_{0},{\bf p}) \neq 0$, 
the spectral densities are regular and given by
\begin{subequations}
\bea
\hat{\rr}^{\,00}(p_{0},{\bf p})& =& \frac{1}{\pi}
\frac{{\rm Im}\,\hat{\Pi}^{00}(p_{0},{\bf p})}{
[p^{2}-{\rm Re}\,\hat{\Pi}^{00}(p_{0},{\bf p})]^{2}+
[{\rm Im}\,\hat{\Pi}^{00}(p_{0},{\bf p})]^{2}}\;,\\
\rr^{t}(p_{0},{\bf p})& =& \frac{1}{\pi}
\frac{{\rm Im}\,\Pi^{t}(p_{0},{\bf p})}{
[p^{2}_{0}-p^{2}-{\rm Re}\,\Pi^{t}(p_{0},{\bf p})]^{2}+
[{\rm Im}\,\Pi^{t}(p_{0},{\bf p})]^{2}}\;.
\eea
\end{subequations}
If  ${\rm Im}\,\hat{\Pi}^{00}(p_{0},{\bf p})$ or
${\rm Im}\,\Pi^t(p_{0},{\bf p})$ vanish, the corresponding
spectral density has a simple pole given by
\be
[p^{2}-{\rm Re}\,\hat{\Pi}^{00}(p_{0},{\bf p})]_{p_{0}=\omega^{00}({\bf p})}=0
\ee
for longitudinal gluons and
\be
[p^{2}_{0}-p^{2}-{\rm Re}\,\Pi^{t}(p_{0},{\bf p})]_{p_{0}=\omega^{t}({\bf p})}=0
\ee
for transverse gluons.

\begin{figure}[ht]
\begin{center}
\vspace*{1cm}
\includegraphics[width=15cm]{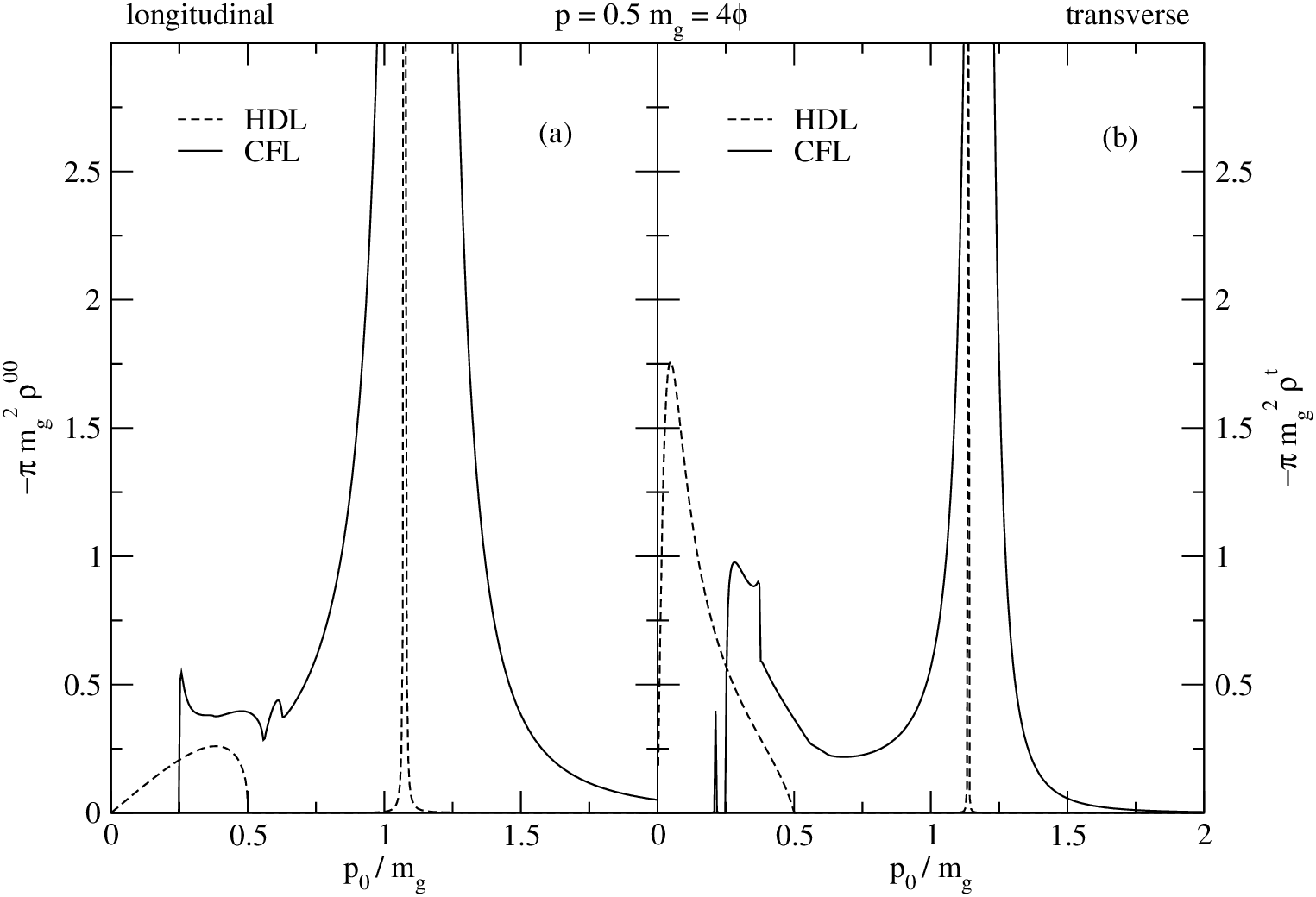}
\caption
{The spectral densities for (a) longitudinal and (b) transverse
gluons for a gluon momentum $p=m_g/2$, with $m_g = 8\, \phi$. 
The dashed lines correspond to the HDL limit.}
\label{figspecdens}
\end{center}
\end{figure}

\begin{figure}[ht]
\begin{center}
\vspace{1cm}
\includegraphics[width=15cm]{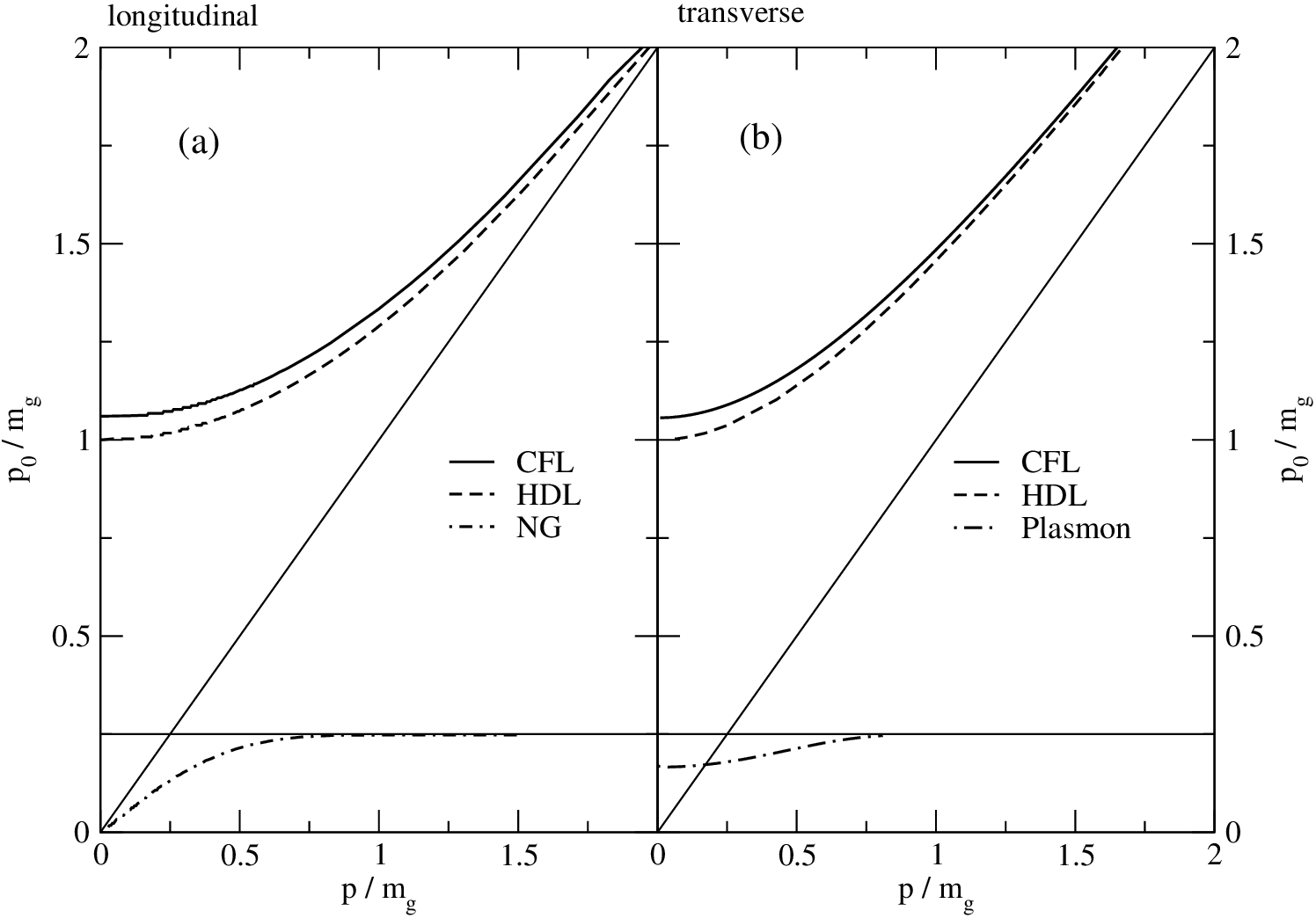}
\caption
{The dispersion relations for (a) longitudinal and (b) transverse
excitations in the CFL phase. The full lines correspond to the
regular longitudinal and transverse excitations. The
dashed lines are for the HDL limit. The dash-dotted line
in part (a) shows the dispersion relation for the Nambu-Goldstone
excitation. The light plasmon dispersion relation
is shown by the dash-dotted line in part (b). As in
Fig.\ \ref{figspecdens}, the value of the gap is chosen such that
$m_g = 8 \, \phi$.}
\label{figplasmon}
\end{center}
\end{figure}

In Fig.\ \ref{figspecdens} we show the spectral densities for
longitudinal and transverse gluons in the CFL phase in comparison
to the HDL limit. Note that there is a delta function-like peak
in the transverse spectral density at an energy $p_0 \simeq 0.21\,
m_g$. This peak corresponds to a collective excitation, the so-called
``light plasmon'' predicted in Ref.\ \cite{plasmon} (see also
\cite{casalbuoni}).
We show the dispersion relation of this collective mode 
in Fig.\ \ref{figplasmon} (b). The
mass $m_{\rm coll} \simeq 1.35\, \phi$ is roughly in agreement
with the value $m_{\rm coll} \simeq 1.362\, \phi$ of Ref.\
\cite{plasmon}. As the momentum increases, the energy of the 
light plasmon excitation approaches $2\, \phi$ from below. 
For momenta larger than $\sim 8\, \phi$, the location of
this excitation branch becomes numerically indistinguishable from
the continuum in the spectral density
above $p_0 = 2\, \phi$, cf.\ Fig.\ \ref{figspecdens}.
Close inspection reveals that the dispersion relation of the
light plasmon has a minimum at a nonzero
value of $p \simeq 1.33\,\phi$, indicating a van Hove singularity.

In Fig.\ \ref{figplasmon} we also show the dispersion relations
for the ``regular'' longitudinal and transverse excitations, as well
as for the Nambu-Goldstone excitation defined by the root
of $P^\mu \, \Pi_{\mu \nu}(P)\, P^{\nu} = 0$ \cite{dirkigor,zarembo}.
For our choice of gauge the
gluon propagator is 4-transverse and this mode does not
mix with the longitudinal
component of the gauge field \cite{dirkigor}. 
Therefore, the Nambu-Goldstone
mode does not appear as a peak in the
longitudinal spectral density, cf.\ Fig.\ \ref{figspecdens}.
We finally note that other collective excitations have been
investigated in Ref.\ \cite{iida}.

\vspace{0.5cm}

In conclusion, we have computed the gluon self-energy 
in the CFL phase as a function of energy and momentum. While the
imaginary parts of the gluon self-energy could be expressed
analytically in terms of elliptic functions (see appendix), the real parts
had to be computed numerically with the help of dispersion
integrals. From 
the real and imaginary parts we constructed the spectral
densities. We confirmed the existence of a low-energy collective
excitation, the so-called ``light plasmon'' predicted in Ref.\
\cite{plasmon}.

\section*{Acknowledgments}
The authors thank Igor Shovkovy for many fruitful discussions.
H.M.\ thanks the Frankfurt International Graduate School of Science
for support. 

\appendix

\section{Calculation of the gluon self-energy}

In this appendix, we compute the individual components
of the gluon self-energy.
According to Eq.\ (31a) of Ref.\ \cite{meissner3},
\begin{subequations}
\bea
\Pi^{00}(P)&=& -\,\frac{g^{2}}{12}\int\frac{d^{3}{\bf k}}{(2\pi)^{3}}
\sum_{e_{1},e_{2}=\pm}(1+e_{1}e_{2}\,{\hat{{\bf k}}}_{1}
\cdot{\hat{{\bf k}}_{2}})
\nonumber \\
&\times &\left[\hspace{.1cm}\left(\frac{1}{p_{0}+{\hat{\e}}_{1}+\e_{2}}-
\frac{1}{p_{0}-{\hat{\e}}_{1}-\e_{2}}\right)
(1-{\hat{N}}_{1}-N_{2})\hspace{.1cm}
\frac{{\hat{\e}}_{1}\e_{2}-\xi_{1}\xi_{2}-
{\hat{\phi}}_{1}\phi_{2}}{2\,{\hat{\e}}_{1}\e_{2}}
\nonumber \right. \\
&+&\hspace{.25cm}\left(\frac{1}{p_{0}+\e_{1}+{\hat{\e}}_{2}}-
\frac{1}{p_{0}-\e_{1}-
{\hat{\e}}_{2}}\right)(1-N_{1}-{\hat{N}}_{2})\hspace{.1cm}
\frac{\e_{1}{\hat{\e}}_{2}-\xi_{1}\xi_{2}-
\phi_{1}{\hat{\phi}}_{2}}{2\,\e_{1}{\hat{\e}}_{2}}
\nonumber \\
&+&7\left(\frac{1}{p_{0}+\e_{1}+\e_{2}}-\frac{1}{p_{0}-\e_{1}-
\e_{2}}\right)(1-N_{1}-N_{2})\hspace{.1cm}\frac{\e_{1}\e_{2}-
\xi_{1}\xi_{2}-2\phi_{1}\phi_{2}/7}{2\,\e_{1}\e_{2}}
\nonumber\\
&+&\hspace{.25cm}\left(\frac{1}{p_{0}+{\hat{\e}}_{1}+\e_{2}}
-\frac{1}{p_{0}-
{\hat{\e}}_{1}-\e_{2}}\right)({\hat{N}}_{1}-N_{2})\hspace{.1cm}
\frac{{\hat{\e}}_{1}\e_{2}+\xi_{1}\xi_{2}+
{\hat{\phi}}_{1}\phi_{2}}{2\,{\hat{\e}}_{1}\e_{2}}
\nonumber \\
&+&\hspace{.25cm}\left(\frac{1}{p_{0}+\e_{1}+{\hat{\e}}_{2}}
-\frac{1}{p_{0}-
\e_{1}-{\hat{\e}}_{2}}\right)(N_{1}-{\hat{N}}_{2})\hspace{.1cm}
\frac{\e_{1}{\hat{\e}}_{2}+\xi_{1}\xi_{2}+
\phi_{1}{\hat{\phi}}_{2}}{2\,\e_{1}{\hat{\e}}_{2}}
\nonumber \\
&+&\left.7\left(\frac{1}{p_{0}+\e_{1}+\e_{2}}-\frac{1}{p_{0}-\e_{1}-
\e_{2}}\right)(N_{1}-N_{2})\hspace{.1cm}\frac{\e_{1}\e_{2}+
\xi_{1}\xi_{2}+2\phi_{1}\phi_{2}/7}{2\,\e_{1}\e_{2}}\right]\; .
\eea
Furthermore, from Eq.\ (31b) of
Ref.\ \cite{meissner3} we obtain
\bea
\Pi^t(P)&=& -\,\frac{g^{2}}{12}\int\frac{d^{3}{\bf k}}{(2\pi)^{3}}
\sum_{e_{1},e_{2}=\pm} (1-e_{1}e_{2}\,\hat{\bf
k}_{1} \cdot \hat{\bf p}\; \hat{\bf k}_{2}  \cdot \hat{\bf p})
\nonumber\\
&\times &\left[\hspace{.1cm}\left(\frac{1}{p_{0}+{\hat{\e}}_{1}+\e_{2}}-
\frac{1}{p_{0}-{\hat{\e}}_{1}-\e_{2}}\right)(1-{\hat{N}}_{1}-N_{2})
\hspace{.1cm}\frac{{\hat{\e}}_{1}\e_{2}-\xi_{1}\xi_{2}+
{\hat{\phi}}_{1}\phi_{2}}{2\,{\hat{\e}}_{1}\e_{2}}
\nonumber \right. \\
&+&\hspace{.25cm}\left(\frac{1}{p_{0}+\e_{1}+{\hat{\e}}_{2}}-
\frac{1}{p_{0}-\e_{1}-
{\hat{\e}}_{2}}\right)(1-N_{1}-{\hat{N}}_{2})\hspace{.1cm}
\frac{\e_{1}{\hat{\e}}_{2}-\xi_{1}\xi_{2}+
\phi_{1}{\hat{\phi}}_{2}}{2\,\e_{1}{\hat{\e}}_{2}}
\nonumber \\ 
&+&7\left(\frac{1}{p_{0}+\e_{1}+\e_{2}}-\frac{1}{p_{0}-\e_{1}
-\e_{2}}\right)(1-N_{1}-N_{2})\hspace{.1cm}\frac{\e_{1}\e_{2}
-\xi_{1}\xi_{2}+2\phi_{1}\phi_{2}/7}{2\,\e_{1}\e_{2}}
\nonumber\\
&+&\hspace{.25cm}\left(\frac{1}{p_{0}+{\hat{\e}}_{1}+\e_{2}}
-\frac{1}{p_{0}-
{\hat{\e}}_{1}-\e_{2}}\right)({\hat{N}}_{1}-N_{2})\hspace{.1cm}
\frac{{\hat{\e}}_{1}\e_{2}+\xi_{1}\xi_{2}-{\hat{\phi}}_{1}
\phi_{2}}{2\,{\hat{\e}}_{1}\e_{2}}
\nonumber \\
&+&\hspace{.25cm}\left(\frac{1}{p_{0}+\e_{1}+{\hat{\e}}_{2}}
-\frac{1}{p_{0}-
\e_{1}-{\hat{\e}}_{2}}\right)(N_{1}-{\hat{N}}_{2})\hspace{.1cm}
\frac{\e_{1}{\hat{\e}}_{2}+\xi_{1}\xi_{2}-
\phi_{1}{\hat{\phi}}_{2}}{2\,\e_{1}{\hat{\e}}_{2}}
\nonumber \\ 
&+&\left. 7\left(\frac{1}{p_{0}+\e_{1}+\e_{2}}-\frac{1}{p_{0}
-\e_{1}-\e_{2}}\right)
(N_{1}-N_{2})\hspace{.1cm}\frac{\e_{1}\e_{2}+\xi_{1}\xi_{2}-
2\phi_{1}\phi_{2}/7}{2\,\e_{1}\e_{2}}\hspace{.1cm}\right]\;,
\eea
and
\bea  
\Pi^\ell(P)&=& -\,\frac{g^{2}}{12}\int\frac{d^{3}{\bf k}}{(2\pi)^{3}}
\sum_{e_{1},e_{2}=\pm} (1-e_{1}e_{2} \hat{\bf k}_{1} \cdot 
\hat{\bf k}_{2}+ 2\, e_{1}e_{2}\,
\hat{\bf k}_1 \cdot \hat{\bf p} \; \hat{\bf k}_2 \cdot \hat{\bf p} )
\nonumber\\
&\times &\left[\hspace{.1cm}\left(\frac{1}{p_{0}+{\hat{\e}}_{1}+\e_{2}}-
\frac{1}{p_{0}-{\hat{\e}}_{1}-\e_{2}}\right)(1-{\hat{N}}_{1}-N_{2})
\hspace{.1cm}\frac{{\hat{\e}}_{1}\e_{2}-\xi_{1}\xi_{2}+
{\hat{\phi}}_{1}\phi_{2}}{2\,{\hat{\e}}_{1}\e_{2}}
\nonumber \right. \\
&+&\hspace{.25cm}\left(\frac{1}{p_{0}+\e_{1}+{\hat{\e}}_{2}}
-\frac{1}{p_{0}-\e_{1}-
{\hat{\e}}_{2}}\right)(1-N_{1}-{\hat{N}}_{2})\hspace{.1cm}
\frac{\e_{1}{\hat{\e}}_{2}-\xi_{1}\xi_{2}+
\phi_{1}{\hat{\phi}}_{2}}{2\,\e_{1}{\hat{\e}}_{2}}
\nonumber \\ 
&+&7\left(\frac{1}{p_{0}+\e_{1}+\e_{2}}-\frac{1}{p_{0}-\e_{1}
-\e_{2}}\right)(1-N_{1}-N_{2})\hspace{.1cm}\frac{\e_{1}\e_{2}
-\xi_{1}\xi_{2}+2\phi_{1}\phi_{2}/7}{2\,\e_{1}\e_{2}}
\nonumber\\
&+&\hspace{.25cm}\left(\frac{1}{p_{0}+{\hat{\e}}_{1}+\e_{2}}-\frac{1}{p_{0}-
{\hat{\e}}_{1}-\e_{2}}\right)({\hat{N}}_{1}-N_{2})\hspace{.1cm}
\frac{{\hat{\e}}_{1}\e_{2}+\xi_{1}\xi_{2}-{\hat{\phi}}_{1}
\phi_{2}}{2\,{\hat{\e}}_{1}\e_{2}}
\nonumber \\
&+&\hspace{.25cm}\left(\frac{1}{p_{0}+\e_{1}+{\hat{\e}}_{2}}-\frac{1}{p_{0}-
\e_{1}-{\hat{\e}}_{2}}\right)(N_{1}-{\hat{N}}_{2})\hspace{.1cm}
\frac{\e_{1}{\hat{\e}}_{2}+\xi_{1}\xi_{2}-
\phi_{1}{\hat{\phi}}_{2}}{2\,\e_{1}{\hat{\e}}_{2}}
\nonumber \\ 
&+&\left. 7\left(\frac{1}{p_{0}+\e_{1}+\e_{2}}-\frac{1}{p_{0}
-\e_{1}-\e_{2}}\right)
(N_{1}-N_{2})\hspace{.1cm}\frac{\e_{1}\e_{2}+\xi_{1}\xi_{2}-
2\phi_{1}\phi_{2}/7}{2\,\e_{1}\e_{2}}\hspace{.1cm}\right]\;.
\eea
The projection $\Pi^{0i}(P)\, \hat{p}_i$ was not explicitly
given in Ref.\ \cite{meissner3}. Starting with Eq.\ (24)
of Ref.\ \cite{meissner3} and following similar steps as
for the other components, we find
\bea
\Pi^{0i}(P)\, \hat{p}_i
&=& \frac{g^{2}}{12}\int\frac{d^{3}{\bf k}}{(2\pi)^{3}}
\sum_{e_{1},e_{2}=\pm} (e_{1} {\hat {\bf k}}_1 \cdot {\hat {\bf p}}
 + e_{2}\, {\hat {\bf k}}_2\cdot {\hat {\bf p}})
\nonumber\\
&\times &\left[\hspace{.1cm}\left(\frac{1}{p_{0}+{\hat{\e}}_{1}+\e_{2}}+
\frac{1}{p_{0}-{\hat{\e}}_{1}-\e_{2}}\right)(1-{\hat{N}}_{1}-N_{2})
\hspace{.1cm}\frac{{\hat{\e}}_{1}\xi_{2}-\xi_{1}\e_{2}}
{2\,{\hat{\e}}_{1}\e_{2}}
\nonumber \right. \\
&+&\hspace{.25cm}\left(\frac{1}{p_{0}+\e_{1}+{\hat{\e}}_{2}}+\frac{1}{p_{0}-\e_{1}-
{\hat{\e}}_{2}}\right)(1-N_{1}-{\hat{N}}_{2})\hspace{.1cm}
\frac{\e_{1}\xi_{2}-\xi_{1}\hat {\e}_{2}}{2\,\e_{1}{\hat{\e}}_{2}}
\nonumber \\
&+&7\left(\frac{1}{p_{0}+\e_{1}+\e_{2}}+\frac{1}{p_{0}-\e_{1}
-\e_{2}}\right)(1-N_{1}-N_{2})\hspace{.1cm}\frac{\e_{1}\xi_{2}
-\xi_{1}\e_{2}}{2\,\e_{1}\e_{2}}
\nonumber\\
&+&\hspace{.25cm}\left(\frac{1}{p_{0}+{\hat{\e}}_{1}+\e_{2}}+\frac{1}{p_{0}-
{\hat{\e}}_{1}-\e_{2}}\right)({\hat{N}}_{1}-N_{2})\hspace{.1cm}
\frac{{\hat{\e}}_{1}\xi_{2}+\xi_{1}\epsilon_{2}}{2\,{\hat{\e}}_{1}\e_{2}}
\nonumber \\
&+&\hspace{.25cm}\left(\frac{1}{p_{0}+\e_{1}+{\hat{\e}}_{2}}+\frac{1}{p_{0}-
\e_{1}-{\hat{\e}}_{2}}\right)(N_{1}-{\hat{N}}_{2})\hspace{.1cm}
\frac{\epsilon_{1}\xi_{2}+ \xi_1{\hat{\e}}_{2}}{2\,\e_{1}{\hat{\e}}_{2}}
\nonumber \\
&+&\left. 7\left(\frac{1}{p_{0}+\e_{1}+\e_{2}}+\frac{1}{p_{0}
-\e_{1}-\e_{2}}\right)
(N_{1}-N_{2})\hspace{.1cm}\frac{\e_{1}\xi_{2}+\xi_{1}\epsilon_{2}
}{2\,\e_{1}\e_{2}}\hspace{.1cm}\right]\;.
\eea
\end{subequations}
Here $\xi_{i}\equiv e_{i}k_{i}-\mu$,
${\bf k}_{1}\equiv {\bf k}$ and ${\bf k}_{2}\equiv {\bf k} - {\bf p}$,
where ${\bf k}$ is the quark three-momentum and ${\bf p}$ is the gluon
three-momentum.
The octet and singlet gap functions for quasiparticles
$(e_{i}=+1)$ and quasiantiparticles $(e_{i}=-1)$
are $\phi_{i}\equiv \phi^{e_{i}}_{\bf 8}$
and ${\hat\phi_{i}}\equiv{\hat \phi}^{e_{i}}_{\bf 1}$, respectively,
and the corresponding excitation energies are
$\e_{i}\equiv\sqrt{\xi^{2}_{i}+\phi^{2}_{i}}$
and $\hat{\e}_{i}\equiv\sqrt{\xi^{2}_{i}+\hat{\phi}^{2}_{i}}$.
The thermal distribution functions are $N_{i}\equiv [\exp(\e_{i}/T)+1]^{-1}$,
and $\hat{N}_{i}\equiv [\exp(\hat{\e}_{i}/T)+1]^{-1}$, respectively.
In the limit $T \rightarrow 0$,
the latter vanish, since $\e_i, \hat{\e}_i > 0$. 
Then the equations simplify to
\begin{subequations} \label{comps}
\bea
\Pi^{00}(P)&=& -\,\frac{g^{2}}{12}\int\frac{d^{3}{\bf k}}{(2\pi)^{3}}
\sum_{e_{1},e_{2}=\pm}(1+e_{1}e_{2}{\hat{{\bf k}}}_{1} \cdot
{\hat{{\bf k}}_{2}})
\nonumber \\ 
&\times &\left[\hspace{.1cm}\left(\frac{1}{p_{0}+{\hat{\e}}_{1}+\e_{2}}-
\frac{1}{p_{0}-{\hat{\e}}_{1}-\e_{2}}\right)\hspace{.1cm}\frac{{\hat{\e}}
_{1}\e_{2}-\xi_{1}\xi_{2}-{\hat{\phi}}_{1}\phi_{2}}{2\,{\hat{\e}}_{1}
\e_{2}}
\nonumber \right. \\
&+&\hspace{.3cm}\left(\frac{1}{p_{0}+\e_{1}+{\hat{\e}}_{2}}-\frac{1}{p_{0}
-\e_{1}- {\hat{\e}}_{2}}\right)\hspace{.1cm}\frac{\e_{1}{\hat{\e}}_{2}-\xi_{1}
\xi_{2}-\phi_{1}{\hat{\phi}}_{2}}{2\,\e_{1}{\hat{\e}}_{2}}
\nonumber \\
&+&\left.7\left(\frac{1}{p_{0}+\e_{1}+\e_{2}}-\frac{1}{p_{0}-\e_{1}-\e_{2}}
\right)
\hspace{.1cm}\frac{\e_{1}\e_{2}-\xi_{1}\xi_{2}-2\phi_{1}\phi_{2}/7}
{2\,\e_{1}\e_{2}}\hspace{.1cm}\right]\label{asd}\;,
\eea
\bea\label{aaa}
\Pi^t(P)&=& -\,\frac{g^{2}}{12}\int\frac{d^{3}{\bf k}}{(2\pi)^{3}}
\sum_{e_{1},e_{2}=\pm} (1-e_{1}e_{2}\hat{\bf k}_{1} \cdot \hat{\bf
p}\; \hat{\bf k}_{2} \cdot \hat{\bf p} )
\nonumber \\
&\times &\left[\hspace{.1cm}\left(\frac{1}{p_{0}+{\hat{\e}}_{1}+\e_{2}}
-\frac{1}{p_{0}-{\hat{\e}}_{1}-\e_{2}}\right)\hspace{.1cm}\frac{{\hat{\e}}
_{1}\e_{2}-\xi_{1}\xi_{2}+{\hat{\phi}}_{1}\phi_{2}}{2\,{\hat{\e}}_{1}
\e_{2}}
\nonumber \right. \\
&+&\hspace{.3cm}\left(\frac{1}{p_{0}+\e_{1}+{\hat{\e}}_{2}}-\frac{1}{p_{0}
-\e_{1}- {\hat{\e}}_{2}}\right)\hspace{.1cm}\frac{\e_{1}{\hat{\e}}_{2}-\xi_{1}
\xi_{2}+\phi_{1}{\hat{\phi}}_{2}}{2\,\e_{1}{\hat{\e}}_{2}}
\nonumber \\
&+&\left.7\left(\frac{1}{p_{0}+\e_{1}+\e_{2}}-\frac{1}{p_{0}-\e_{1}-\e_{2}}
\right)
\hspace{.1cm}\frac{\e_{1}\e_{2}-\xi_{1}\xi_{2}+2\phi_{1}\phi_{2}/7}
{2\,\e_{1}\e_{2}}\hspace{.1cm}\right]\;,
\eea
\bea\label{abc}
\Pi^\ell(P)&=& -\,\frac{g^{2}}{12}\int\frac{d^{3}{\bf k}}{(2\pi)^{3}}
\sum_{e_{1},e_{2}=\pm} (1 - e_{1}e_{2}\hat{\bf k}_{1} \cdot \hat{\bf
k}_2 + 2\, e_{1}e_{2}\hat{\bf k}_{1} \cdot \hat{\bf
p}\; \hat{\bf k}_{2} \cdot \hat{\bf p} )
\nonumber \\
&\times &\left[\hspace{.1cm}\left(\frac{1}{p_{0}+{\hat{\e}}_{1}+\e_{2}}
-\frac{1}{p_{0}-{\hat{\e}}_{1}-\e_{2}}\right)\hspace{.1cm}\frac{{\hat{\e}}
_{1}\e_{2}-\xi_{1}\xi_{2}+{\hat{\phi}}_{1}\phi_{2}}{2\,{\hat{\e}}_{1}
\e_{2}}
\nonumber \right. \\
&+&\hspace{.3cm}\left(\frac{1}{p_{0}+\e_{1}+{\hat{\e}}_{2}}-\frac{1}{p_{0}
-\e_{1}- {\hat{\e}}_{2}}\right)\hspace{.1cm}\frac{\e_{1}{\hat{\e}}_{2}-\xi_{1}
\xi_{2}+\phi_{1}{\hat{\phi}}_{2}}{2\,\e_{1}{\hat{\e}}_{2}}
\nonumber \\
&+&\left.7\left(\frac{1}{p_{0}+\e_{1}+\e_{2}}-\frac{1}{p_{0}-\e_{1}-\e_{2}}
\right)
\hspace{.1cm}\frac{\e_{1}\e_{2}-\xi_{1}\xi_{2}+2\phi_{1}\phi_{2}/7}
{2\,\e_{1}\e_{2}}\hspace{.1cm}\right]\;,
\eea
\bea\label{xyz}
\Pi^{0i}(P)\, \hat{p}_i
&=& \frac{g^{2}}{12}\int\frac{d^{3}{\bf k}}{(2\pi)^{3}}
\sum_{e_{1},e_{2}=\pm} (e_{1}\, {\hat {\bf k}}_1\cdot{\hat {\bf p}}
 + e_{2}\,{\hat {\bf k}}_2 \cdot {\hat {\bf p}})
\nonumber\\
&\times &\left[\hspace{.1cm}\left(\frac{1}{p_{0}+{\hat{\e}}_{1}+\e_{2}}+
\frac{1}{p_{0}-{\hat{\e}}_{1}-\e_{2}}\right)
\hspace{.1cm}\frac{{\hat{\e}}_{1}\xi_{2}-\xi_{1}\e_{2}}
{2\,{\hat{\e}}_{1}\e_{2}}
\nonumber \right. \\
&+&\hspace{.25cm}\left(\frac{1}{p_{0}+\e_{1}+{\hat{\e}}_{2}}
+\frac{1}{p_{0}-\e_{1}-
{\hat{\e}}_{2}}\right)\hspace{.1cm}
\frac{\e_{1}\xi_{2}-\xi_{1}\hat {\e}_{2}}{2\,\e_{1}{\hat{\e}}_{2}}
\nonumber \\
&+&7\left.\left(\frac{1}{p_{0}+\e_{1}+\e_{2}}+\frac{1}{p_{0}-\e_{1}
-\e_{2}}\right)\hspace{.1cm}\frac{\e_{1}\xi_{2}
-\xi_{1}\e_{2}}{2\,\e_{1}\e_{2}}\right]\;.
\eea

\end{subequations}

In the following, we evaluate the imaginary
part of $\Pi^{00}$ explicitly;
the calculation of the other components is similar.

\subsection {Imaginary parts}

We are interested in the retarded self-energy, so we analytically
continue $p_0 \rightarrow p_0 + i \eta$ in Eqs.\ (\ref{comps}).
Then, using the Dirac identity 
\be\label{ddd}
\frac {1}{x+i\eta}={\cal P}\frac{1}{x}-i\pi\delta(x)\;,
\ee
where ${\cal P}$ stands for the principal-value prescription, 
we find 
\bea\label{cdf}
{\rm Im}\,\Pi^{00}(P)&=& \pi\,\frac{g^{2}}{12}\int\frac{d^{3}{\bf k}}
{(2\pi)^{3}}\sum_{e_{1},e_{2}=\pm}(1+e_{1}e_{2}{\hat{{\bf k}}}_{1}
\cdot {\hat{{\bf k}}_{2}})
\nonumber \\
&\times &\left\{\hspace{.1cm}\bigg[\hspace{.1cm}\delta\big(p_{0}+
{\hat{\e}}_{1}+\e_{2}\big)-
\delta\big(p_{0}-{\hat{\e}}_{1}-\e_{2}\big) \bigg]\,
\frac{{\hat{\e}}_{1}\e_{2}-
\xi_{1}\xi_{2}-{\hat{\phi}}_{1}\phi_{2}}{2\,{\hat{\e}}_{1}\e_{2}}
\nonumber \right. \\
&+&\hspace{.4cm}\bigg[\hspace{.1cm}\delta\big(p_{0}+\e_{1}+{\hat{\e}}_{2}\big)
-\delta\big(p_{0}-\e_{1}-
{\hat{\e}}_{2}\big)\bigg]\,\frac{\e_{1}{\hat{\e}}_{2}-\xi_{1}\xi_{2}-
\phi_{1}{\hat{\phi}}_{2}}{2\,\e_{1}{\hat{\e}}_{2}}
\nonumber \\
&+&\hspace{.4cm}\left. \bigg[\hspace{.1cm}\delta\big(p_{0}+\e_{1}+\e_{2}\big)-
\delta\big(p_{0}-\e_{1}-\e_{2}\big)\bigg]\,
\frac{7(\e_{1}\e_{2}-\xi_{1}\xi_{2})-2\phi_{1}\phi_{2}}
{2\,\e_{1}\e_{2}}
\right\}
\eea

Since Eq.\ (\ref{cdf}) is an odd function of $p_{0}$, ${\rm Im}\,\Pi(-p_{0}
,{\bf p})\equiv -{\rm Im}\,\Pi(p_{0},{\bf p})$, the calculation can be
restricted to positive values of the energy $p_{0}\ge 0$. Hence,
the first delta-function in each square bracket in Eq.\ (\ref{cdf})
can be dropped. 

Moreover, we are interested in
gluon energies and momenta $p_{0}, p \ll\mu$. Consequently,
in order to have a vanishing argument of the remaining
delta-functions, the quasiparticle energies should not be too large either, 
$\e_i, \hat{\e}_{i} \ll \mu$. 
Then, only the terms with $e_{1}=e_{2}=+1$
will contribute; for either $e_{1}$ or $e_{2}=-1$, $\e_i \simeq \hat{\e}_{i} 
\simeq |k_{i}+\mu|\sim\mu$ is too far from the Fermi surface
to make a contribution for $p_0 \ll \mu$. 
Shifting the integration variable so that
${\bf k}_{1}={\bf k}+{\bf p}/2$ and ${\bf k}_{2}={\bf k}-{\bf p}/2$,
we find using $p \ll \mu$
\be \label{A5}
{\hat{\bf k}_{1}} \cdot {\hat{\bf k}_{2}} \simeq  1\;,\;\;\;\;
\xi_{1,2} \simeq  \xi \pm \frac{{\bf p} \cdot {\hat{\bf k}}}{2}
\equiv\xi_{\pm}\;, 
\ee
where we have defined $\xi\equiv k-\mu$.
Furthermore, we denote
\be
\e_{\pm} \equiv\sqrt{\xi^{2}_{\pm}+\phi^{2}} \;, \;\;\;\;
{\hat{\e}}_{\pm}\equiv\sqrt{\xi^{2}_{\pm}+{\hat{\phi}}^{2}}\;.
\ee

Setting $\phi_{1}\simeq\phi_{2}\equiv\phi$, ${\hat{\phi}}_{1}
\simeq{\hat{\phi}}_{2}\equiv{\hat{\phi}} \simeq 2 \phi$ 
in weak coupling, we obtain
\bea\label{bbb}
{\rm Im}\,\Pi^{00}(P)= -\,\frac{g^{2}\pi}{6}\int \frac{d^{3}{\bf k}}
{(2\pi)^{3}}&&\hspace{-.5cm}\bigg[\hspace{.1cm}\delta\big(p_{0}-
{\hat{\e}}_{+}-\e_{-}\big)\,\frac{{\hat{\e}}_{+}\e_{-}-\xi_{+}
\xi_{-}-2\phi^{2}}{2{\hat{\e}}_{+}\e_{-}}
\nonumber \\
&+&\hspace{-.1cm}\delta\big(p_{0}-\e_{+}-{\hat{\e}}_{-}\big)\,
\frac{\e_{+}{\hat{\e}}_{-}
-\xi_{+}\xi_{-}-2\phi^{2}}{2\e_{+}{\hat{\e}}_{-}}
\nonumber \\ 
&+&\hspace{-.1cm}\delta\big(p_{0}-\e_{+}-\e_{-}\big)\,\frac{7(\e_{+}
\e_{-}-\xi_{+}\xi_{-})-2\phi^{2}}{2\e_{+}\e_{-}}\hspace{.1cm}
\bigg]\;.
\eea

Choosing ${\bf p}=(0,0,p)$ the
integration over the polar angle $\varphi$ becomes trivial.
The integral over $\xi$ can be performed using the delta-functions.
Denoting $x= \cos \theta$, the roots of the arguments of the
delta-functions in Eq.\ (\ref{bbb}) are (in the order of
appearance)
\begin{subequations}
\bea
\xi^{*}_{1,2}(x)&=&\frac{-3px\phi^{2}\pm p_{0}\sqrt{(p^{2}x^{2}
-p^{2}_{0}+9\phi^{2})(p^{2}x^{2}-p^{2}_{0}+\phi^{2})}}
{2(p^{2}x^{2}-p^{2}_{0})}\;, \\
\xi^{*}_{3,4}(x)&=&\frac{3px\phi^{2}\pm p_{0}\sqrt{(p^{2}x^{2}
-p^{2}_{0}+9\phi^{2})(p^{2}x^{2}-p^{2}_{0}+\phi^{2})}}
{2(p^{2}x^{2}-p^{2}_{0})}\;, \\
\xi^{*}_{5,6}(x)&=&\pm\frac{p_{0}}{2}\sqrt{1-\frac{4\phi^{2}}
{p^{2}_{0}-p^{2}x^{2}}}\; .
\eea
\end{subequations}
Then,
\bea\label{fff}
{\rm Im}\,\Pi^{00}(P)& = &-\pi m^{2}_{g}\, \frac{\phi^2}{3\,p\,p_0}\,
\Bigg\{\Theta\big(p_0-3\phi\big)
\int_0^{u_1} \,dy\, \Big[\,\frac{9}{\,\sqrt{(y^{2}-1+9\Phi^{2})
(y^{2}-1+\Phi^{2})}\,}
\nonumber \\
& & - \,\frac{10+9\Phi^{2}}{\,(1-y^{2})\sqrt{(y^{2}-1+9\Phi^{2})
(y^{2}-1+\Phi^{2})}\,}
 + \frac{18\Phi^{2}}{\,(1-y^{2})^{2}\hspace{.1cm}
\sqrt{(y^{2}-1+9\Phi^{2})(y^{2}-1+\Phi^{2})}\,}\Big]
\nonumber \\
& & + \,\Theta\big(p_0-2\phi\big)\int_0^{u_2} dy\,\,\frac{5+9y^{2}}
{\,(y^{2}-1)^{3/2}\hspace{.1cm}\sqrt{y^{2}-1+4\Phi^{2}}\,}
\Bigg\}\;,
\eea
where $y\equiv p\,x/p_{0}$, $\Phi\equiv\phi/p_{0}$,
$u_1={\rm min}(p/p_{0},\sqrt{1-9\phi^{2}/p^{2}_{0}\,\,}\,)$,
$u_2={\rm min}(p/p_{0},\sqrt{1-4\phi^{2}/p^{2}_{0}\,\,}\,)$ and 
$m_g$ is the gluon mass parameter (squared), $m_g^2=N_f g^2\mu^2 /
6\pi^2$, for $N_f = 3$.

Using the elliptic integrals of first, second, and third kind,
\begin{subequations}
\bea
F(\varphi,k)& = &\int^{\varphi}_{0}\frac{d\alpha}{\sqrt{1-k^{2}
\sin^{2}\alpha}}\;, \\
E(\varphi,k) & = & \int^{\varphi}_{0}d\alpha\,\sqrt{1-k^{2}
\sin^{2}\alpha}\;, \\
\Pi(\varphi,l,k) & = &  \int^{\varphi}_{0}\frac{d\alpha}{
(1+l\,\sin^{2}\alpha)\sqrt{1-k^{2}\sin^{2}\alpha}}\;,
\eea
\end{subequations}
and the complete elliptic integrals
of the first, ${\rm \bf{K}}(k)=F(\pi/2,k)$, the second kind, ${\rm \bf{E}}(k)
=E(\pi/2,k)$ and the third kind ${\rm \bf{\Pi}}\big(l,r\big) =
\Pi\big(\pi/2,l,r\big)$
in Eq.\ (\ref{fff}), we obtain the final result
\bea
{\rm Im}\,\Pi^{00}(P)&=&\frac{\pi m^{2}_{g}}{6}\frac{p_0}{p}\Bigg(
\Theta\big(p_{0}-3\phi\big)\,
\frac{s^{2}}{\sqrt{4-s^2}}
\Bigg\{\,\Theta\big(E_{p}^{\bf 18}-p_0\big)\bigg[
\hspace{.1cm}9\,{\rm \bf {K}}(t')-\bigg(10+\frac{9}{4}\,s^{2}\bigg)
\,{\rm \bf{\Pi}}\big(l,t'\big)
\nonumber\\
&-&\frac{9}{2}\,s^{2}
\frac{d}{dn}\frac{1}{n}{\rm \bf{\Pi}}\big(\frac{l}{n},t'\big)
\bigg|_{n=1}\hspace{.1cm}\bigg]
+\Theta\big(p_0 - E_{p}^{\bf 18}\big)\bigg[\,9\,F(\alpha{'}\,,t')-
\bigg(10+\frac{9}{4}\,s^{2}\bigg)\,\Pi\big(\alpha{'},l,t'\big)
\nonumber\\
&-&\frac{9}{2}\,s^{2}\frac{d}{dn}\frac{1}{n}\,
\Pi\big(\alpha{'},\frac{l}{n},t'\big)\bigg|_{n=1}\hspace{.1cm}\bigg]\Bigg\}
- \Theta\big(p_{0}-2\phi\big)\Bigg\{\Theta\big(E_{p}^{\bf 88}-p_0\big)
\bigg[\,7\,{\rm \bf{E}}(t)-\frac{9}{2}\,s^{2}\,{\rm \bf{K}}(t)\,\bigg]
\nonumber\\
&+& \Theta\big(p_0 - E_{p}^{\bf 88}\big)\bigg[\,7\,E(\alpha,t)-
\frac{9}{2}\,s^{2}F(\alpha,t)-7\,\frac{p}{p_{0}}
\sqrt{1-\frac{4\phi^{2}}{p^{2}_{0}-p^{2}}\,\,}\,\,\bigg]\Bigg\}\Bigg)\;,
\label{A11}
\eea
where $t'=\sqrt{(p^{2}_{0}-9\phi^{2})/(p^{2}_{0}-\phi^{2})\,\,}$,
$t=\sqrt{1-4\phi^{2}/p^{2}_{0}\,\,}$, $\alpha'={\rm arcsin}
[p/\sqrt{p^{2}_{0}-9\phi^{2}\,\,}]$,
$\alpha={\rm arcsin}[p/(tp_0)]$, $l=-1+9\phi^{2}/p^{2}_{0}$ and $s=2\phi/p_0$.

The imaginary parts of the other components of the gluon self-energy
can be obtained analogously from Eqs.\ (\ref{aaa}),
(\ref{abc}), and (\ref{xyz}). In addition to Eq.\ (\ref{A5})
we employ
\be
{\hat {\bf k}}_{1} \cdot {\hat {\bf p}}\;{\hat{\bf k}}_{2}
\cdot{\hat {\bf p}} \simeq ({\hat {\bf k}}\cdot{\hat {\bf p}})^{2}\;,\;\;\;
{\hat {\bf k}}_{1} \cdot {\hat {\bf p}} + {\hat{\bf k}}_{2}
\cdot{\hat {\bf p}} \simeq 2\,{\hat {\bf k}}\cdot{\hat {\bf p}}\;,
\ee
and find
\begin{subequations} \label{A13}
\bea
\hspace{-.6cm}{\rm Im}\,\Pi^{t}(P)&=&\frac{\pi m^{2}_{g}}{12}\frac{p_0}{p}\,
\left[\frac{s^2}{\sqrt{4-s^{2}\,}\,}\hspace{.1cm} \Theta\big(p_0-3\,\phi\big)
\Bigg(\,\Theta\big(E_{p}^{\bf 18}-p_0\big)
\bigg\{\hspace{.1cm} \frac{p_0^2}{p^{2}}\Big(1-\frac{s^{2}}{4}\Big)
\,{\rm \bf{E}}(t')+\Big[1-\frac{p_0^2}{p^{2}}\big(11+2s^2\big)\,\Big]\right.
\nonumber\\
&\times&\,{\rm \bf{K}}(t')-\Big[10\big(1-\frac{p_0^2}{p^{2}}\big)+
\frac{9s^2}{4}\,\big(1-3\frac{p_0^2}{p^{2}}\big)\Big]
\,{\rm \bf {\Pi}}(l,t')-\frac{9s^2}{2}
\Big(1-\frac{p_0^{2}}{p^2}\Big)\frac{d}{dn}\frac{1}{n}{\rm \bf {\Pi}}\,(
\frac{l}{n},t')\bigg|_{n=1}\bigg\}
\nonumber\\
&+&\Theta\big(p_{0} - E_{p}^{\bf 18}\big)\bigg\{\hspace{.1cm}
\frac{p_0^2}{p^{2}}\Big(1-\frac{s^{2}}{4}\Big)\,E(\alpha',t')
+\Big[1-\frac{p_0^2}{p^{2}}\big(11+2s^2\big)\,\Big]\,F(\alpha',t')
-\Big[10\big(1-\frac{p_0^2}{p^{2}}\big)
\nonumber\\
&+&\frac{9s^2}{4}\,\big(1-3\frac{p_0^2}{p^{2}}\big)\Big]\,
\Pi(\alpha',l,t')-\frac{9s^2}{2}\,
\Big(1-\frac{p_0^{2}}{p^2}\Big)\frac{d}{dn}\frac{1}{n}\Pi\,(\alpha',
\frac{l}{n},t')\bigg|_{n=1}
\bigg\}\Bigg)
+\Theta(p_{0}-2\phi)
\nonumber \\
&\times&\Bigg(\Theta\big(E_{p}^{\bf 88}-p_0\big)
\bigg\{\hspace{.1cm}
\Big[\frac{5p_0^2}{2p^{2}}\,s^{2}-7\big(1-\frac{p_0^2}{p^2}\big)\Big]{\rm \bf{E}}(t)+
\frac{s^2}{2}\,\Big(5 - 19\frac{p_0^2}{p^2}\,\Big)\,{\rm \bf {K}}(t)\bigg\}
+ \Theta\big(p_{0} - E_{p}^{\bf 88}\big)
\nonumber\\
&\times&\bigg\{\hspace{.1cm}
\Big[\frac{5p_0^2}{2p^{2}}\,s^{2}-7\big(1-\frac{p_0^2}{p^2}\big)\Big]E(\alpha,t)
+\frac{s^2}{2}\,\Big(5 - 19\frac{p_0^2}{p^2}\Big)F(\alpha,t)+\left.7\frac{p}{p_{0}}
\Big(1-\frac{p_0^2}{p^2}\Big)\sqrt{1-\frac{4\phi^{2}}
{p^{2}-p^{2}_{0}}}\hspace{.1cm} \bigg\}\hspace{.1cm} \Bigg)\right]\,,
\eea

\bea
{\rm Im}\,\Pi^{\ell}(P)&=&-\frac{\pi m^{2}_{g}}{6}\,
\frac{p_0^3}{p^3}\,\Bigg(
\hspace{.1cm}\frac{s^2}{\sqrt{4-s^2}} \Theta\big(p_0-3\,\phi\big)
\Bigg\{\Theta\big(E_{p}^{\bf 18}-p_0\big)
\bigg[\hspace{.1cm}\Big(1-\frac{s^2}{4}\Big){\rm \bf{E}}(t')+\Big
(10+\frac{27}{4}\,s^2\Big)
{\rm \bf{\Pi}}\big(l,t'\big)
\nonumber\\
&-&\Big(11+2s^2\Big){\rm\bf{K}}(t')+\frac{9}{2}s^2\,\frac{d}{dn}\,
\frac{1}{n}\,{\rm\bf{\Pi}}\big(l,t'\big)
\bigg|_{n=1}\bigg]\,+\,\Theta\big(p_{0} - E_{p}^{\bf 18}\big)
\bigg[\hspace{.1cm}\Big(1-\frac{s^2}{4}\Big)E(\alpha',t')
\nonumber \\
&+&\Big(10+\frac{27}{4}\,s^2\Big)\Pi\big(\alpha',l,t'\big)-
\Big(11+2s^2\Big)F(\alpha',t')
+\frac{9}{2}s^2\frac{d}{dn}\,\frac{1}{n}\,
\Pi\big(\alpha',l,t'\big)\bigg|_{n=1}\bigg]\Bigg\}
\nonumber \\
&+&\Theta(p_{0}-2\phi)\Bigg\{\Theta\big(E_{p}^{\bf 88}-p_0\big)
\bigg[\hspace{.1cm}\Big(7+\frac{5}{2}\,s^2\Big){\rm\bf{E}}(t)
-\frac{19}{2}\,s^2\,{\rm\bf{K}}(t)\bigg]
\nonumber\\
&+& \Theta\big(p_{0} - E_{p}^{\bf 88}\big)
\bigg[\hspace{.1cm}\Big(7+\frac{5}{2}\,s^2\Big)E(\alpha,t)-
\frac{19}{2}\,s^2\,F(\alpha,t)-\frac{7\,p}{p_0}
\sqrt{1-\frac{4\,\phi^2}{p_0^2-p^2}\,}\hspace{.1cm} \bigg]\hspace{.1cm}
\Bigg\}\Bigg)\;,
\eea

\bea
{\rm Im}\,\Pi^{0i}(P)\,{\hat p_i}&=&-\frac{\pi m^{2}_{g}}
{6}\,\frac{p_0^2}{p^2}\,\Bigg(
\hspace{.1cm} \frac{2s^2}{\sqrt{4-s^2\,}}\,\Theta\big(p_0-3\,\phi\big)
\Bigg\{\Theta\big(E_{p}^{\bf 18}-p_0\big)
\bigg[\hspace{.1cm}5\,{\rm\bf{K}}(t')-(5+\frac{9}{4}\,s^2)\,
{\rm\bf{\Pi}}\big(l,t'\big)
\nonumber\\
&-&\frac{9}{4}\,s^2\,\frac{d}{dn}\,\frac{1}{n}
\,{\rm\bf{\Pi}}\big(l,t'\big)\bigg|_{n=1}\bigg]
\,+\,\Theta\big(p_0 - E_{p}^{\bf 18}\big)
\bigg[\hspace{.1cm}5\,F(\alpha',t')-(5+\frac{9}{4}\,s^2)\,
\Pi\big(\alpha',l,t'\big)
\nonumber\\
&-&\frac{9}{4}\,s^2\,\frac{d}{dn}\,\frac{1}{n}\,
\Pi\big(\alpha',l,t'\big)\bigg|_{n=1}\bigg]\Bigg\}
- 7\,\Theta(p_{0}-2\phi)\Bigg\{\Theta\big(E_{p}^{\bf 88}-p_0\big)
\bigg[\hspace{.1cm}{\rm\bf{E}}(t)-s^2\,{\rm\bf{K}}(t)\bigg]
\nonumber\\
&+& \Theta\big(p_{0} - E_{p}^{\bf 88}\big)
\bigg[\hspace{.1cm}{\rm\bf{E}}(\alpha,t)
-\frac{p}{p_0}\sqrt{1-\frac{4\,\phi^2}{p_0^2-p^2}\,}
-s^2\,F(\alpha,t)\hspace{.1cm} \bigg]\hspace{.1cm}
 \Bigg\}\Bigg)\;.
\eea

\end{subequations}

In the  limit $\phi\to 0$, we reproduce the
HDL limit,
\begin{subequations}
\bea\label{ccc}
\lim\limits_{\phi\to 0}{\rm Im}\, \Pi^{00}(P)& \equiv&
{\rm Im}\,\Pi^{00}_{0}(P)\;,\\
\lim\limits_{\phi\to 0}{\rm Im}\, \Pi^{t}(P)&\equiv&
{\rm Im}\, \Pi^{t}_{0}(P)\;,\\
\lim\limits_{\phi\to 0}{\rm Im}\, \Pi^{\ell}(P)&\equiv&
{\rm Im}\,\Pi^{\ell}_{0}(P)\;,\\
\lim\limits_{\phi\to 0}{\rm Im}\, \Pi^{0i}(P)\,{\hat p}_i&\equiv&
{\rm Im}\, \Pi^{0i}_{0}(P)\, {\hat p}_i\;.
\eea
\end{subequations}

\subsection{Real parts}

There are two possibilities to compute the real parts of the gluon self-energy.
Either, one evaluates a principal-value integral, cf.\
Eq.\ (\ref{ddd}), or one employs the dispersion integral
\be\label{disp}
{\rm Re}\hspace{.1cm}\Pi(p_{0},{\bf p})\equiv
\frac{1}{\pi}{\cal P}\int_{-\infty}^{\infty}
d\omega\, \frac{{\rm Im}\,\Pi(\omega,
{\bf p})}{\omega-p_{0}}+C\;,
\ee
where $C$ is a (subtraction) constant. If ${\rm Im}\,\Pi(\omega,
{\bf p})$ is an odd function of $\omega$, ${\rm Im}\,\Pi(-\omega,
{\bf p})\equiv - {\rm Im}\,\Pi(\omega ,{\bf p})$ the dispersion
integral becomes
\be
{\rm Re}\hspace{.1cm}\Pi(p_{0},{\bf p})\equiv \frac{1}{\pi}\, {\cal P}
\int_{0}^{\infty}d\omega\, {\rm Im}\,\Pi_{\rm odd}(\omega,{\bf p})\Big(\frac{1}
{\omega+p_{0}}+\frac{1}{\omega-p_{0}}\Big)+C\;,
\ee
and if it is an even function of $\omega$, ${\rm Im}\,\Pi(-\omega,
{\bf p})\equiv + {\rm Im}\,\Pi(\omega,
{\bf p})$ we have
\be
{\rm Re}\hspace{.1cm}\Pi(p_{0},{\bf p})\equiv \frac{1}{\pi}\, {\cal P}
\int_{0}^{\infty}d\omega\, {\rm Im}\,\Pi_{\rm even}(\omega ,{\bf p})
\Big(\frac{1}
{\omega-p_{0}}-\frac{1}{\omega+p_{0}}\Big)+C\;,
\ee
where in both cases $\Pi(p_{0},{\bf p})$
is assumed to be analytic in the upper complex $p_{0}$ plane.

The values of the constants $C^{00},\, C^t,\, C^\ell,$ and $C^{0i}$ are
determined by the large-$p_0$ dependence of the self-energy.
Thus, it does not matter which color-superconducting phase
we consider, and the constants assume the same values as
for the 2SC phase,
$C^{00}=C^{0i}=0,\,C^t = C^\ell = m_g^2$, cf.\ Ref.\ \cite{dirkigor}.

\end{document}